\documentclass[showpacs,showkeys]{revtex4}
\usepackage{graphicx}
\usepackage{amsmath}

\newcommand{\be}{\begin{equation}}
\newcommand{\ee}{\end{equation}}
\newcommand{\bea}{\begin{eqnarray}}
\newcommand{\eea}{\end{eqnarray}}

\begin{document}
\title{ 
QED at a finite chemical potential
}
\date{\today}

\author{R. Narayanan}
\email{rajamani.narayanan@fiu.edu}
\affiliation{Department of Physics, Florida International University, Miami,
FL 33199.}

\begin{abstract}
We consider multi-flavor QED on a finite lattice at a finite chemical
potential and
show that the partition function only depends on the variables,
$\left(\frac{\mu_i}{q_i} -
  \frac{\mu_1}{q_1}\right)$,
for $i=2,\cdots$, where $q_i$, $i=1\cdots$ are integer valued charges
of the various flavors and $\mu_i$, $i=1\cdots$ are the dimensionless
chemical potentials of the various flavors.
\end{abstract}

\pacs{12.20.-m}
\keywords{QED, chemical potential}
\maketitle

Consider multi-flavor QED regularized on the lattice. We will assume
that the continuum theory is defined by first defining the theory on a
finite periodic
lattice and then taking the limit where the extent in all directions
goes off to infinity. We will not be concerned in this paper as to
whether such a theory
has
a well defined continuum limit. For definiteness, we could think about
two dimensional multi-flavor QED which has a well defined continuum limit.

Let $U_\nu({\bf n})$ be the link variable connecting the sites ${\bf n}$ and
${\bf n}+\hat \nu$ on a $d$ dimensional periodic lattice, $L_1\times
\cdots L_d$. Consider the class of gauge fields given by
$U_\nu({\bf n}) e^{i\frac{2\pi h_\nu}{L_\nu}}$ with $0\le h_\nu < 1$
for $\nu=1,\cdots d$.
Gauge fields with different choices of the dimensionless variables, $h_\nu$, within this class are
not gauge equivalent but have the same gauge action.
The fermion determinant, on the other hand, depends on
the variables, $h_\nu$, which we will refer to as the toron variables.
Consider a fermion with integer charge $q_i$ and a dimensionless chemical potential $\mu_i$.
The chemical potential is introduced~\cite{Hasenfratz:1983ba} by multiplying the parallel
transporter in the forward $\nu=d$ direction by $e^{\frac{2\pi \mu_i}{L_d}}$ 
and in the backward $\nu=d$ direction by
$e^{-\frac{2\pi\mu_i}{L_d}}$. The factor of $2\pi$ is introduced for
convenience and we keep the dimensionless chemical
potential, $\mu_i$, fixed as we take $L_d\to\infty$.

The fermionic determinant is a function of
$U_\nu({\bf n})$, $\nu=1,\cdots,d$; $h_\nu$ for 
$\nu=1,\cdots, d-1$; and $z_i=h_d - i\frac{\mu_i}{q_i}$.
Since the gauge action does not depend on the toron variables, we can integrate
the fermion determinant over these variables using the uniform measure. Consider, for
simplicity,
a lattice fermion operator of the na\"ive, Wilson or staggered type.
We will show later that our arguments will also apply to the overlap
Dirac operator under certain mild assumptions.
The lattice fermion operator is a finite matrix on a finite
lattice. Focussing on the dependence on $h_d$ alone, we see that the
fermion determinant is a finite polynomial in $e^{i\frac{2\pi q_iz_i}{L_d}}$ and
$e^{-i\frac{2\pi q_iz_i}{L_d}}$. The fermion determinant will be an analytic
function of $z_i$ in the complex plane. Since the fermion determinant
is gauge invariant, it will be periodic under $z_i\to z_i+1$.
A contour integral in the complex plane results in the integral over
$h_d$ in the range $[0,1]$ to be independent of $\mu_i$. If we consider a theory with many
flavors,
that all have the same value for $\frac{\mu_i}{q_i}$, then again the
integral
over $h_d$ will yield a result that is independent of all the $\mu_i$.
In other words, a multi-flavor theory of QED at a finite chemical
potential
can only depend on the variables, 
$\left(\frac{\mu_i}{q_i} -
  \frac{\mu_1}{q_1}\right)$,
for $i=2,\cdots$. This result will also hold for an anomaly free
chiral QED as long as the lattice formulation is gauge invariant, at
least in the continuum limit.
This is the main observation in this paper.

\begin{figure}
\centerline{\includegraphics[scale=0.75]{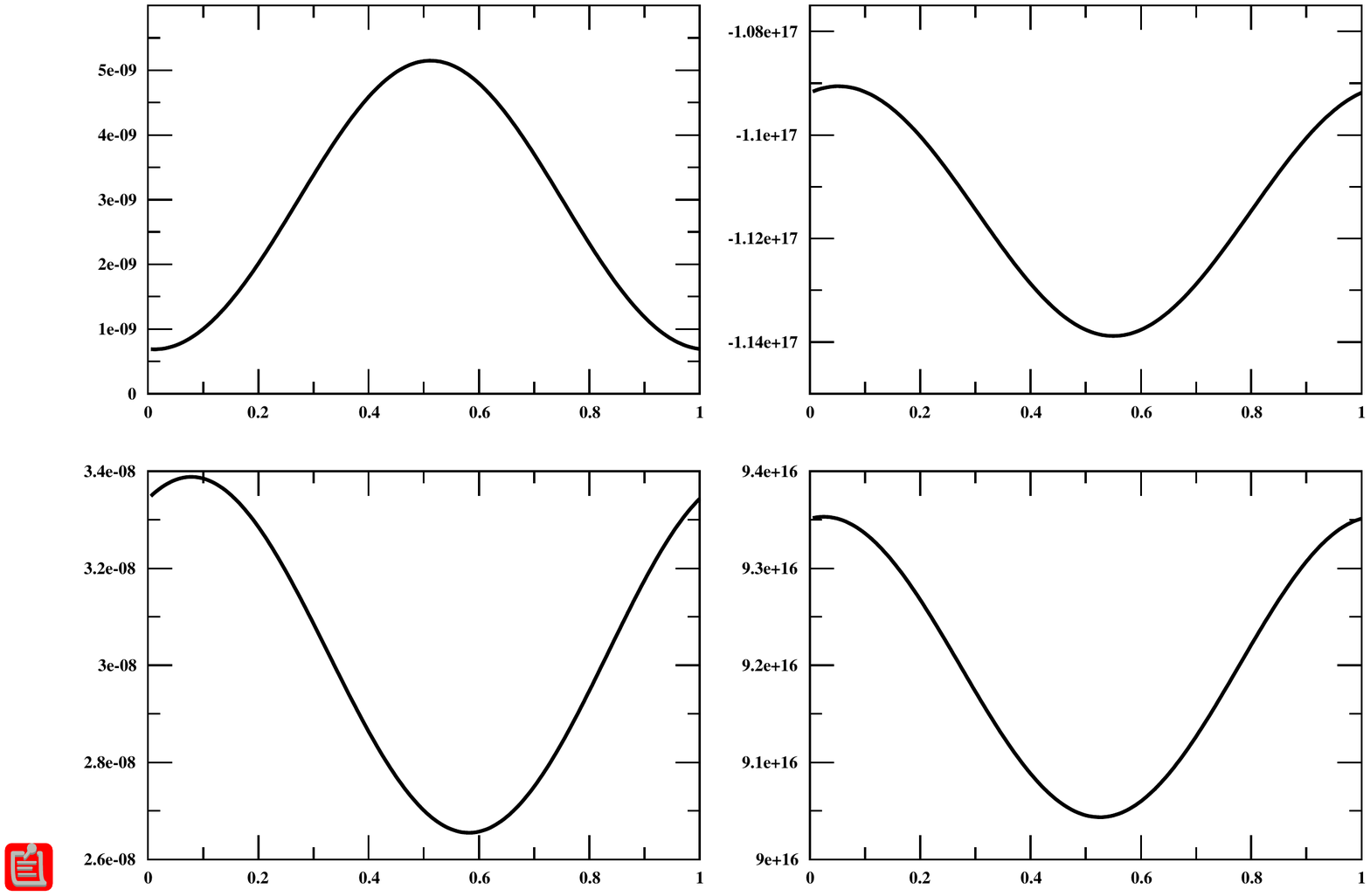}}
\caption{\label{fig1}
The left top and bottom panels show the behavior of the
  determinant of the overlap Dirac operator in the zero and unit
  topological sectors.
The right top and bottom panels show the behavior of the
  determinant of the Wilson Dirac operator in the zero and unit
  topological sectors. The results are on a $7\times 7$ lattice and
the dimensionless chemical potential is set to $0$ and the $x$ axis
shows the value of $h_2$.
}
\end{figure}

\begin{figure}
\centerline{\includegraphics[scale=0.75]{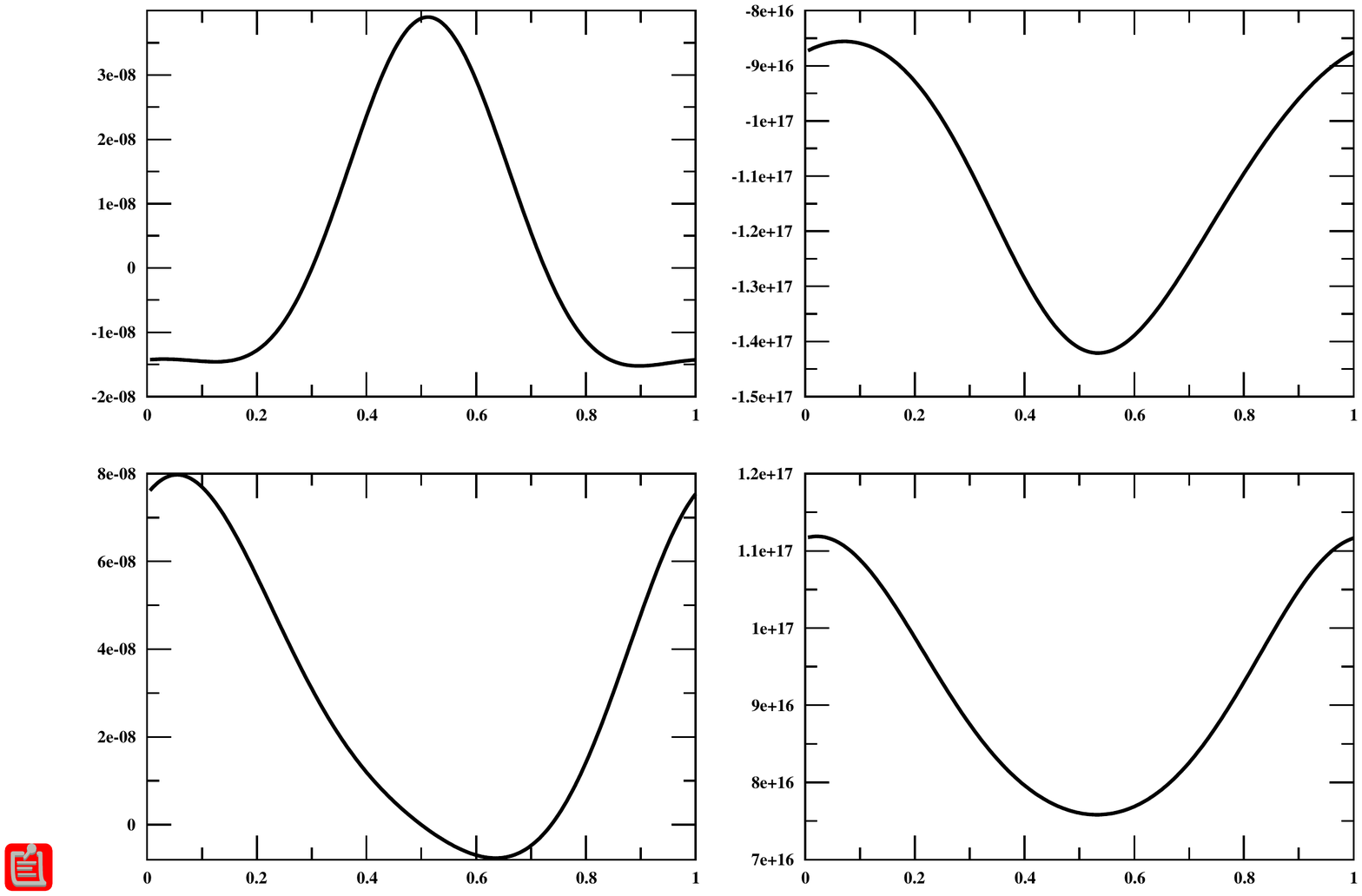}}
\caption{\label{fig2}
The left top and bottom panels show the behavior of the
  determinant of the overlap Dirac operator in the zero and unit
  topological sectors.
The right top and bottom panels show the behavior of the
  determinant of the Wilson Dirac operator in the zero and unit
  topological sectors. The results are on a $7\times 7$ lattice and
the dimensionless chemical potential is set to $0.5$ and the $x$ axis shows the value
of $h_2$.}
\end{figure}

\begin{figure}
\centerline{\includegraphics[scale=0.75]{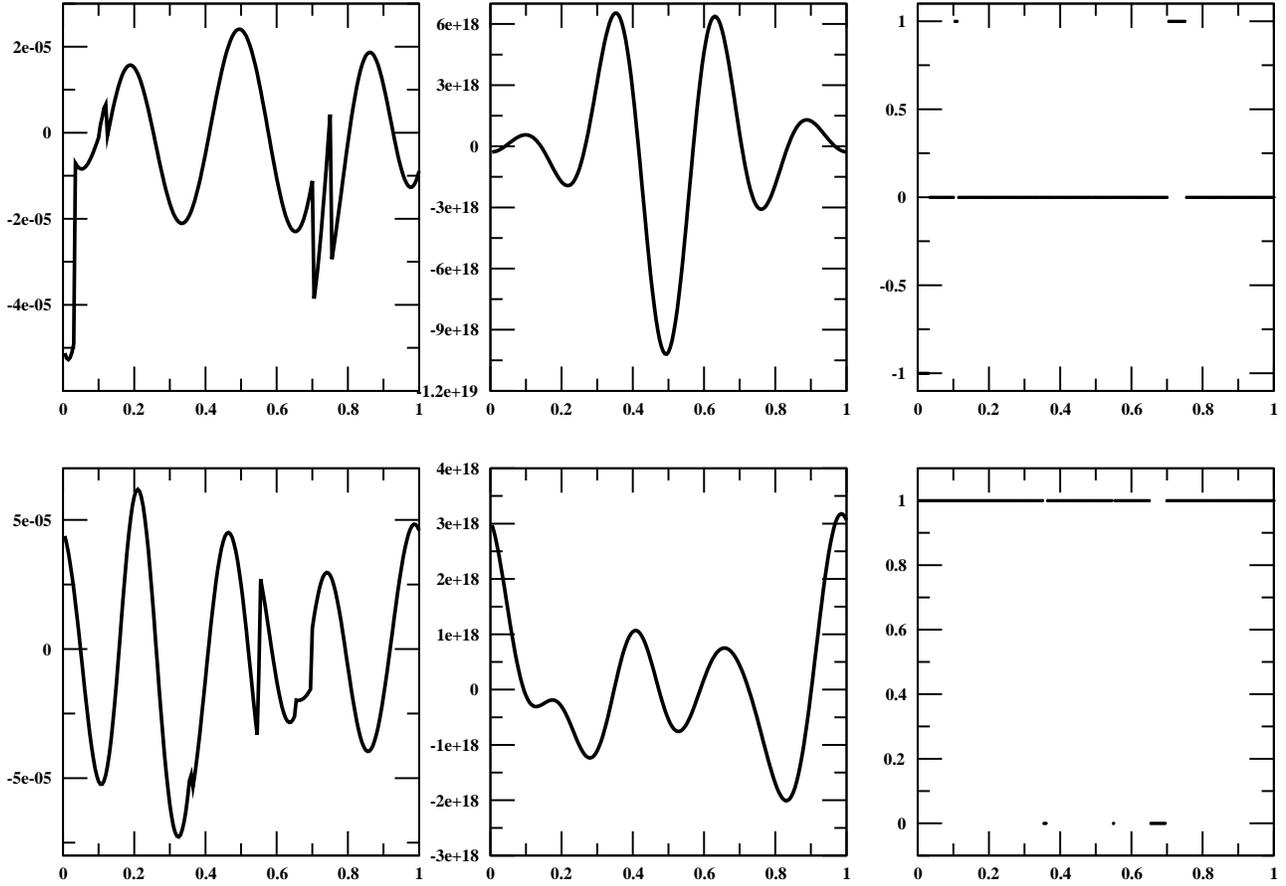}}
\caption{\label{fig3}
The left top and bottom panels show the behavior of the
  determinant of the overlap Dirac operator in the zero and unit
  topological sectors with zero chemical potential.
The middle top and bottom panels show the behavior of the
  determinant of the Wilson Dirac operator in the zero and unit
  topological sectors. The right top and bottom panels shows the {\sl
    topological charge} measured  by the overlap Dirac operator in the
  presence of a chemical potential for the presumed zero and unit
  topological charge configuration.
The results are on a $7\times 7$ lattice and
the dimensionless chemical potential is set to $1.0$ and the $x$ axis shows the value of $h_2$.
} 
\end{figure}

\begin{figure}
\centerline{\includegraphics[scale=0.75]{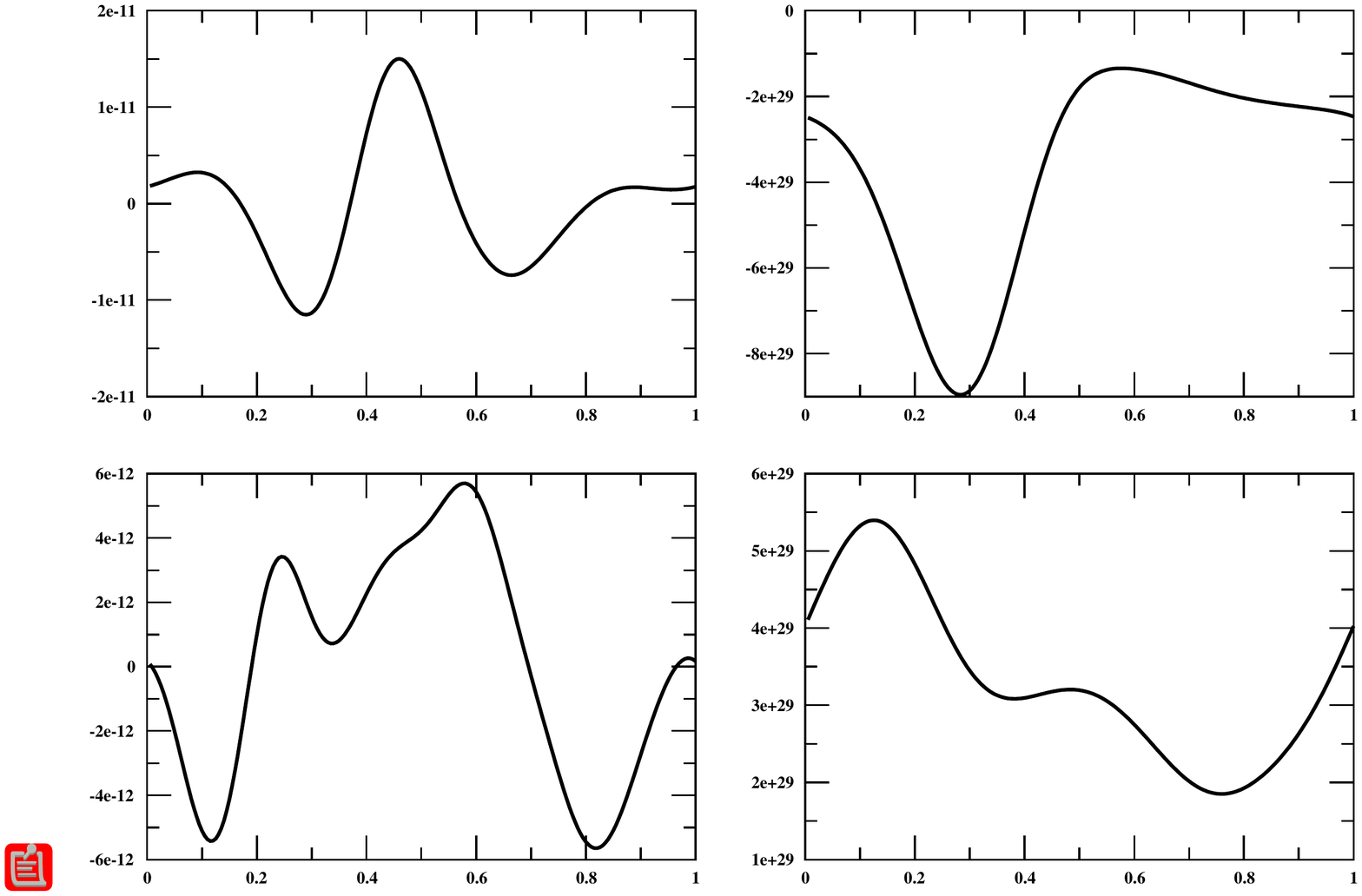}}
\caption{\label{fig4}
The left top and bottom panels show the behavior of the
  determinant of the overlap Dirac operator in the zero and unit
  topological sectors with zero chemical potential.
The right top and bottom panels show the behavior of the
  determinant of the Wilson Dirac operator in the zero and unit
  topological sectors. 
The results are on a $9\times 9$ lattice and
the dimensionless chemical potential is set to $1.0$ and the $x$ axis shows the value of $h_2$.
} 
\end{figure}

Massless Schwinger model in the presence of a chemical potential was
first studied in~\cite{Fischler:1978ms}. The problem was treated in
the Hamiltonian formalism. In order to deal with a finite problem, a
uniform charge background was introduced in a finite region of space.
This causes an explicit breaking of translational invariance. The
ground state 
is a classical Wigner crystal which is not destroyed by quantum
fluctuations.
An explicit chemical potential term was introduced in the Hamiltonian
formalism in~\cite{Kao:1994wv}. Their formalism also had to break
translational invariance and they conclude that there is a
inhomogeneous chiral condensate in the Schwinger model at finite
density.
A path integral formulation of the problem, again with the
introduction of a chemical potential that breaks translational
invariance results in an inhomogenous chiral condensate~\cite{Christiansen:1996iu}.
References to a possible inhomogenous chiral
condensate in the Schwinger
model are still being discussed in the literature~\cite{Schon:2000qy,Basar:2008ki,Kojo:2009ha}.
The problem is discussed in~\cite{Metlitski:2006id} where the author argues that
the inhomogeneous chiral condensate in the Schwinger model at finite
density
is an artifact of the explicit breaking of translational invariance in the
formalism. 
The generalized Thirring model was analyzed in~\cite{Sachs:1995dm} 
and it has been explicitly shown that the chiral condensate does not 
depend on the chemical potential. 
 Our observation in this paper clearly shows that physics does not depend
on the chemical potential in the Schwinger model.

We now proceed to discuss the case of overlap fermions in some detail
when the chemical potential is introduced as
in~\cite{Bloch:2006cd}. Since
the definition involves the sign function of a complex matrix, it is
not apriori clear if the arguments presented above will apply.
This problem is addressed using a specific numerical example, namely,
the Schwinger model with a finite chemical potential, $\mu$.
Gauge fields are generated at a fixed coupling using the gauge action
described in~\cite{Kikukawa:1997dv} with
zero and unit topological charge. The determinant 
of both the Wilson-Dirac operator and the overlap-Dirac operator was
computed in a
fixed
gauge field background that has zero topological charge. The determinant 
of both the Wilson-Dirac operator and the overlap-Dirac operator (we
exclude the zero mode in this case) was also computed in a fixed
gauge field background that has unit topological charge. 
Averages of these two quantities over all gauge fields for the
overlap-Dirac
operator enter the
computation of the chiral condensate in the massless Schwinger model.

We first consider two sample gauge fields, one with zero topological
charge and one with unit topological charge as set by the gauge
fields, on a $7\times 7$ lattice. The same gauge field 
is used for several different values of the chemical
potential. Plots of the real part of the determinants~\footnote{The
fermion determinant in a  charge conjugated gauge field background is the complex
conjugate of the fermion determinant in the original gauge field background.} for  three
sample values of the chemical potential, namely, $\mu=0,0.5,1.0$ are
shown in Figs.\ref{fig1},\ref{fig2},\ref{fig3} respectively. In all
three cases, the determinant of the Wilson Dirac operator with a
specific choice for the mass term used in the definition of the
overlap Dirac operator kernel behaves smoothly as a function of $h_2$.
The determinants do depend on $h_2$ and the variation
increases as one increases the chemical potential. Furthermore, the
function
even changes sign for $\mu=1.0$. In spite of this, we explicitly
verified that the
integral over $h_2$ (which remains complex for a fixed gauge field
background) is independent of $\mu$ as expected by the
analytical argument presented in the beginning of the paper.

The sign function of the Wilson Dirac operator need not be a smooth
function of $h_2$. This is due to the fact it depends on the sign of
the
real part of the eigenvalue of the Wilson Dirac operator and this can
change
as a function of $h_2$. The trace of the sign function of the Wilson
Dirac operator is defined as twice the topological charge of the gauge
field.
Whereas, the results are consistent with the topological charge of the
constructed
gauge field background for small values of the chemical potential, it
does
not remain consistent for large values of the chemical potential as
seen in the right panel of Fig.~\ref{fig3}. The top panel should have
been consistent with zero and the bottom panel should have been
consistent
with unity. Since this is not the case for $\mu=1$, the determinant of
the overlap Dirac operator in the two different topological sectors is
not a smooth function of $h_2$ as seen in the left panels of
Fig.~\ref{fig3}. As a consequence the independence of the integral of
the
determinant of the overlap Dirac operator on the chemical potential
breaks
down beyond a certain value of the chemical potential. This is due to
the discretization of the dimensional chemical potential, $\mu$, over
a finite number of slices, $L_d$, taken to be $7$ in Fig.~\ref{fig3}.
If we increase the number of slices, $L_d$, keeping the physical gauge coupling
fixed,
we found that the value of the chemical potential where the
independence
breakdown increases. For example, we could only go up to $\mu=0.7$ on
$7\times 7$ lattice at a given coupling but we could go up to $\mu=1.0$
on a $9\times 9$ lattice at the same physical coupling.
This is illustrated for the case of $\mu=1.0$ on a
$9\times 9$ lattice
where the behavior of the determinants of the overlap Dirac operator
are smooth as a function of $h_2$. The problem with the overlap Dirac
operator
in the presence of a chemical potential was anticipated
in~\cite{Bloch:2006cd} but is expected not to affect the continuum limit.

We have presented an analytical argument in this paper that there is
no dependence on the chemical potential in the path
integral formalism of QED. This result is valid
within the lattice formulation and in the continuum limit. Whereas,
there is some justification to the  argument presented
in~\cite{Metlitski:2006id}
toward the problems with breaking translation invariance in the
Hamiltonian formalism, we have shown that the main reason for
recovering
the correct behavior in the presence of a chemical potential is the
integration over the toron variable as emphasized
in~\cite{Sachs:1995dm} and further emphasized in~\cite{Langfeld:2011rh}.
One does not see the toron variable, $h_d$, in the
hamiltonian formalism since one starts in the Coloumb gauge. The
realization of the toron variable in the Hamiltonian formalism would
be
to integrate over all boundary conditions for fermions in the
Euclidean time (temperature) direction~\cite{Gross:1980br}. 

We have not performed an analytic continuation in the chemical
potential in this paper. One could reproduce the central argument in
this paper by working with an imaginary chemical
potential~\footnote{We would like to thank
  Philippe de Frocrand for making us aware of~\cite{Bender:1992gn}.}.
Since there is no periodicity when the chemical potential is real, we
refrained
from using imaginary chemical potential -- our results are valid for
all values of the real chemical potential. A non-trivial dependence on
the chemical potential will be seen in the two flavor Schwinger
model
and finite density phase transitions are expected in four dimensional two flavor QED.

The continuous toron variable becomes a discrete $Z_N$ variable in a $SU(N)$ gauge theory 
and our argument of independence on the chemical potential will not go
through~\footnote{If we analytically continue to imaginary chemical
  potentials, the issue of Roberge-Weiss transitions~\cite{Roberge:1986mm} need to be adressed.}
In the limit of $N\to\infty$, we have a continuous toron variable.
Therefore,
we expect the toron variable to play a part in the analysis of the 't Hooft model in
the
presence of a chemical potential as discussed in~\cite{Galvez:2009rq}.
The Gross-Neveu model~\cite{Schon:2000qy,Basar:2008ki} is different in
this aspect since one does not integrate over all possible
fermionic
boundary conditions in the Euclidean time direction. This will be the
case even if we introduce a bosonic variable to convert the four-fermi
coupling into a fermion bilinear since we will have a Gaussian term
for the bosonic field.

\begin{acknowledgments} 
R.N. acknowledges partial support by the NSF under grant number
PHY-0854744.  
\end{acknowledgments}


\begin{thebibliography}{99}
\bibitem{Hasenfratz:1983ba} 
  P.~Hasenfratz and F.~Karsch,
  Phys.\ Lett.\ B {\bf 125}, 308 (1983).
\bibitem{Fischler:1978ms}
  W.~Fischler, J.~B.~Kogut and L.~Susskind,
  Phys.\ Rev.\ D {\bf 19} (1979) 1188.
\bibitem{Kao:1994wv} 
  Y.~-C.~Kao and Y.~-W.~Lee,
  Phys.\ Rev.\ D {\bf 50}, 1165 (1994).
\bibitem{Christiansen:1996iu} 
  H.~R.~Christiansen and F.~A.~Schaposnik,
  Phys.\ Rev.\ D {\bf 53}, 3260 (1996)
  [hep-th/9602063].
\bibitem{Schon:2000qy} 
  V.~Schon and M.~Thies,
  In *Shifman, M. (ed.): At the frontier of particle physics, vol. 3* 1945-2032
  [hep-th/0008175].
\bibitem{Basar:2008ki} 
  G.~Basar and G.~V.~Dunne,
  Phys.\ Rev.\ D {\bf 78}, 065022 (2008)
  [arXiv:0806.2659 [hep-th]].
\bibitem{Kojo:2009ha} 
  T.~Kojo, Y.~Hidaka, L.~McLerran and R.~D.~Pisarski,
  Nucl.\ Phys.\ A {\bf 843}, 37 (2010)
  [arXiv:0912.3800 [hep-ph]].
\bibitem{Metlitski:2006id} 
  M.~A.~Metlitski,
  Phys.\ Rev.\ D {\bf 75}, 045004 (2007)
  [hep-th/0609046].
\bibitem{Sachs:1995dm} 
  I.~Sachs and A.~Wipf,
  Annals Phys.\  {\bf 249}, 380 (1996)
  [hep-th/9508142].
\bibitem{Bloch:2006cd} 
  J.~C.~R.~Bloch and T.~Wettig,
  Phys.\ Rev.\ Lett.\  {\bf 97}, 012003 (2006)
  [hep-lat/0604020].
\bibitem{Kikukawa:1997dv} 
  Y.~Kikukawa, R.~Narayanan and H.~Neuberger,
  Phys.\ Rev.\ D {\bf 57}, 1233 (1998)
  [hep-lat/9705006].
\bibitem{Langfeld:2011rh} 
  K.~Langfeld and A.~Wipf,
  Annals Phys.\  {\bf 327}, 994 (2012)
  [arXiv:1109.0502 [hep-lat]].
\bibitem{Gross:1980br} 
  D.~J.~Gross, R.~D.~Pisarski and L.~G.~Yaffe,
  Rev.\ Mod.\ Phys.\  {\bf 53}, 43 (1981).
\bibitem{Bender:1992gn} 
  I.~Bender, T.~Hashimoto, F.~Karsch, V.~Linke, A.~Nakamura, M.~Plewnia, I.~O.~Stamatescu and W.~Wetzel,
  Nucl.\ Phys.\ Proc.\ Suppl.\  {\bf 26}, 323 (1992).
\bibitem{Roberge:1986mm} 
  A.~Roberge and N.~Weiss,
  Nucl.\ Phys.\ B {\bf 275}, 734 (1986).
\bibitem{Galvez:2009rq} 
  R.~Galvez, A.~Hietanen and R.~Narayanan,
  Phys.\ Lett.\ B {\bf 672}, 376 (2009)
  [arXiv:0812.3449 [hep-lat]].
\end{thebibliography}
\end{document}